\documentstyle[12pt]{article}
\begin{document}
\large

\begin{center}
{\bf ON THE ANOMALOUS STRUCTURES OF THE VECTOR LEPTONIC CURRENTS}
\end{center}
\vspace{1cm}
\begin{center}
{\bf Rasulkhozha S. Sharafiddinov}
\end{center}
\vspace{1cm}
\begin{center}
{\bf Institute of Nuclear Physics, Uzbekistan Academy of Sciences,
Tashkent, 702132 Ulugbek, Uzbekistan}
\end{center}
\vspace{1cm}

Each of existing types of the electric charges come forwards in the system
as the source of a kind of the dipole moment. Therefore, to investigate
these regularities we have established the compound structures of Dirac
and Pauli form factors. They state that the electron possesses as well
as the anomalous electric charge.

\newpage
Owing to the vector nature of virtual photon, the elastic scattering
of electrons by spinless nuclei depends on the Dirac $F_{1e}(q^{2})$
and Pauli $F_{2e}(q^{2})$ form factors of light leptons \cite{1}. They
are of course the functions of the square of four - dimentional momentum
transfer. However, in spite of large number of works dedicated to the
interaction between the electron and field of emission, thus far remains
many uncertainties both in the structures and in the behavior of these
currents. Usually it is accepted that $F_{2e}(0)$ is equal to the electron
anomalous magnetic moment \cite{2}, and its full magnetic moment is defined
by the combination \cite{3} of form factors \cite{4}
\begin{equation}
\mu_{e}^{full}=\frac{F_{1e}(0)}{2m_{e}}+F_{2e}(0).
\label{1}
\end{equation}

It appears that here $F_{1e}(0)$ gives the electric charge leading to the
appearance of the electron normal magnetic moment.

Such a procedure were based actually on the assumption \cite{5} of that the
functions $F_{1e}(q^{2})$ and $F_{2e}(q^{2})$ are not the Fourier transforms
of the spatial distributions of the electric charge and magnetic moment
of a particle. This is explained by some consequences of the classical
model of an extensive electron \cite{6}.

According to the classical theory of electromagnetic mass \cite{7},
the availability of the eigenenergy $E_{0}$ of the electron electrostatic
field implies the existence of the electric part of the electron
rest mass:
$$m_{e}^{em}=\frac{E_{0}}{c^{2}}.$$

The opinion has been speaked out that all the mass of the electron is equal
to its electromagnetic mass. Such an idea called simply a hypothesis of field
mass and testifies in favor of the unsteadity of charge distribution
of the electron.

We start from the duality of matter that the mass and charge of a particle
correspond to the most diverse form of the same regularity of the nature
of this field \cite{8,9}. It states that each of all possible types of
charges arises as a consequence of the availability of a kind of the
inertial mass \cite{10}. Thereby such a mechanism leads to the appearance
of the intraelectron interratio between the forces of the electric and
unelectric nature. Therefore, the charge distribution of the electron
must be steady.

The purpose of the present work is to discuss some consequences and
implications implied from the above - mentioned regularities of the
nature of matter. They give of course the justification of that in the
same presentation as the form factors $F_{1e}(q^{2})$ and $F_{2e}(q^{2})$
was used are not in the states to explain the observed vector picture of
the electron. For understanding the mechanism of the anomalous interaction
of Pauli at the fundamental level, one must elucidate the compound structures
of these functions.

From such a purpose, we not only must write the form factors
$F_{ie}(q^{2})$ in the form
\begin{equation}
F_{ie}(q^{2})=f_{ie}(0)+A_{ie}(\vec{q^{2}})+...
\label{2}
\end{equation}
but also need conclude that each of existing types of the electric charges
come forwards in the system as the source of a kind of the dipole moment.
Herewith the independent components $f_{ie}(0)$ coincide with the normal
size of the electric charge and magnetic moment of the electron:
\begin{equation}
f_{1e}(0)=e_{e}^{norm}, \, \, \, \,
f_{2e}(0)=\mu_{e}^{norm}=
\frac{e_{e}^{norm}}{2m_{e}^{norm}},
\label{3}
\end{equation}
where and further it is necessary to keep in mind that $e_{e}^{norm}$ for
a particle (antiparticle) has the negative (positive) sign.

The second terms $A_{ie}(\vec{q^{2}})$ characterize the dependence of form
factors on the square of three - dimensional momentum transfer $\vec{q^{2}}$
and at the emission of a real photon ($\vec{q^{2}}$=0) are reduced
to the values
\begin{equation}
A_{1e}(0)=e_{e}^{anom}, \, \, \, \,
A_{2e}(0)=\mu_{e}^{anom}=\frac{e_{e}^{anom}}{2m_{e}^{anom}}.
\label{4}
\end{equation}
Here $m_{e}^{norm}$ and $m_{e}^{anom}$ are the Coulomb normal and anomalous
masses. Insofar as the full electric mass is concerned, we will start from
the fact that the Coulomb mass and charge of a particle correspond to two
form of the same regularity of its electric nature.
Therefore, we conclude \cite{9} that
\begin{equation}
m_{\nu}^{E}=m_{\nu}^{norm}+m_{\nu}^{anom}+....
\label{5}
\end{equation}

So, it is seen that any of form factors $F_{1e}(q^{2})$ and $F_{2e}(q^{2})$
includes in self both normal and anomalous interactions between the electron
and field of emission. In other words, they must be Fourier transforms of the
spatial density of charge and moment. Their value at zero four - dimensional
momentum transfer $(q^{2}=0)$ defines the full static size of the electric
charge and magnetic moment of a particle:
\begin{equation}
F_{1e}(0)=e_{e}^{full}=e_{e}^{norm}+e_{e}^{anom}+...,
\label{6}
\end{equation}
\begin{equation}
F_{2e}(0)=\mu_{e}^{full}=\mu_{e}^{norm}+\mu_{e}^{anom}+....
\label{7}
\end{equation}

By following the compound structures of form factors (\ref{3}), we get
\begin{equation}
f_{2e}(0)=\frac{f_{1e}(0)}{2m_{e}^{norm}}.
\label{8}
\end{equation}

Exactly the same one can found from (\ref{4}) that
\begin{equation}
A_{2e}(0)=\frac{A_{1e}(0)}{2m_{e}^{anom}}.
\label{9}
\end{equation}

According to the presented here point of view, the electron possesses
as well as the anomalous electric charge which has an estimate of
\begin{equation}
e_{e}^{anom}=\frac{\alpha}{2\pi}
\left(\frac{m_{e}^{anom}}{m_{e}^{norm}}\right)e_{e}^{norm}
\label{10}
\end{equation}
in assuming that the size of $A_{2e}(0)$ is equal to the electron
Schwinger magnetic moment:
\begin{equation}
A_{2e}(0)=\mu_{e}^{anom}=
\frac{\alpha}{2\pi}\frac{e_{e}^{norm}}{2m_{e}^{norm}}.
\label{11}
\end{equation}

To brighter reveal our ideas one must apply to the process of elastic
scattering of electrons and their neutrinos by spinless nuclei as to the
source of unique information about structures of leptonic currents.
It is already clear from (\ref{2}) that in the case of one - photon
exchange only the independent components of form factors are responsible
for the interaction with matter. Therefore, a study of the behavior
of light leptons $(l=e,\nu_{e})$ in the nucleus charge field leads
us to the equation \cite{9}
\begin{equation}
2m_{l}^{norm}\frac{f_{2l}(0)}{f_{1l}(0)}=\pm 1.
\label{12}
\end{equation}

Comparison of (\ref{12}) with (\ref{8}) say in favor of correspondence
principle which states that each terms of the expansions (\ref{2}) correspond
to the definite approximations \cite{9}. Under such circumstances the
possibility of the inclusion of the anomalous phenomena $A_{ie}(\vec{q^{2}})$
in the discussion is realized only in the second Born approximation.
Nevertheless, without loss of generality, we must have in view of that any
non - zero component of the interaction of Pauli implies the availability
of a kind of the Dirac interaction. Of course, the above - noted regularities
of vector picture of the electron and its neutrino open up new possibilities
for developments of our sights at the nature of matter.


\newpage
\begin{thebibliography}{99}
\bibitem{1} R.B. Begzhanov and R.S. Sharafiddinov, {\it Mod. Phys.
Lett.} A {\bf 15} (2000) 557; {\it Izv. Russ. Acad. Nauk Ser.
Fiz.} {\bf 64} (2000) 2221
\bibitem{2} J. Schwinger, {\it Phys. Rev.} {\bf 76} (1949) 790
\bibitem{3} B.K. Kerimov, T.R. Aruri and M.Ya. Safin,
{\it Izv. Acad. Nauk SSSR. Ser. Fiz.} {\bf 37} (1973) 1768
\bibitem{4} R.G. Sachs, {\it Phys. Rev.} B {\bf 136} (1962) 281
\bibitem{5} R.G. Schas, {\it Phys. Rev.} {\bf 126} (1962) 2256, Appendix II.
\bibitem{6} E. Fermi, {\it Rend. Lincei} {\bf 31} (1922) 184, 306;
{\it Phys. Zeit.} {\bf 23} (1922) 340
\bibitem{7} E. Fermi, {\it Nuovo Cimento} {\bf 25} (1923) 159
\bibitem{8} R.S. Sharafiddinov, {\it in Proc. Ukrain - Russian Grav. Conf.
"Gravitation, Cosmology and Relativistic Astrophysics"} (November
8-11, 2000, {\it Kharkov, Ukraine}), p.25
\bibitem{9} R.S. Sharafiddinov, {\it Spacetime \& Substance}
{\bf 1} (2000) 176
\bibitem{10} R.S. Sharafiddinov, {\it Spacetime \& Substance}
{\bf 3} (2002) 47
\end{thebibliography}
\end{document}